\documentclass[twocolumn,amssymb, showpacs, amsmath,superscriptaddress,prb]{revtex4-1}
\usepackage{graphicx}
\usepackage{dcolumn}
\usepackage{bm}
\usepackage{xcolor}

\begin{document}

\title{How the vortex lattice of a superconductor becomes disordered:\\ a study by scanning tunneling spectroscopy}

\author {M. Zehetmayer}
\affiliation {Atominstitut, Vienna University of Technology, 1020 Vienna, 
Austria}
 \email{zehetm@ati.ac.at}

\begin{abstract}

Order-disorder transitions take place in many physical systems, but observing them in detail in real materials is difficult. In two- or quasi-two-dimensional systems, the transition has been studied by computer simulations and experimentally in electron sheets, dusty plasmas, colloidal and other systems. Here I show the different stages of defect formation in the vortex lattice of a superconductor while it undergoes an order-disorder transition by presenting real-space images of the lattice from scanning tunneling spectroscopy. When the system evolves from the ordered to the disordered state, the predominant kind of defect changes from dislocation pairs to single dislocations, and finally to defect clusters forming grain boundaries. Correlation functions indicate a hexatic-like state preceding the disordered state. The transition in the microscopic vortex distribution is mirrored by the well-known spectacular second peak effect observed in the macroscopic 
current density of the superconductor.

\end{abstract}

\maketitle

Though studied for many years, the physics of vortex matter of a superconductor  \cite{Bla94a,Bra95a} is far from being unraveled. There are basically three competing energies, namely vortex-vortex repulsion, vortex pinning, and thermal fluctuations, that determine the state of vortex matter \cite{Mik01a,Kie04a}. While vortex-vortex repulsion gives rise to an ordered vortex lattice, vortex pinning and thermal fluctuations make the lattice disordered. Changing the magnetic field, the temperature or the defect structure modifies the balance of the competing energies, which evokes transitions between different vortex matter states. 

The macroscopic critical current of a superconductor may have a second maximum (fishtail effect) at a high magnetic field, which is the result of an 
order-disorder transition of the vortex lattice \cite{Dae90a,Pal00a,Iav08a,Hec14a}. How this transition 
takes place in detail by the formation of lattice defects is a fundamental question that has been addressed 
by theoretical work \cite{Kie00a} and computer simulations \cite{Cha04a,Mor05a} 
thus far. I have addressed this question by experiment, namely by pushing the vortex lattice  from the ordered to the disordered state in a controlled way in small steps and studying the corresponding real-space lattice images acquired by scanning tunneling spectroscopy.

Order-disorder transitions have been investigated in two- or 
quasi-two-dimensional systems by computer simulations \cite{Str88a,Ber11a} and experimentally in electron sheets \cite{Gri79a}, dusty plasmas \cite{Tho96a}, colloidal \cite{Mur87a,Mar96a,Deu13a} and other systems \cite{Str88a,Gru07a}. Most results agree with theory\cite{Kos73a,Nel79a,Hal78a,You79a,Chu83a} predicting that the lattice becomes disordered in a two-stage process including the intermediate hexatic state. The properties of the vortex matter are fundamentally different to these systems, which makes studying its order-disorder transition worthwhile. Indeed, the essential point of this work is to show in detail the different stages of the order-disorder transition of the vortex matter in real-space, driven by the successive generation of dislocation pairs, single dislocations, and finally grain boundaries. Additionally, I will analyze the different states, including a possible hexatic-like state, by evaluating correlation functions and will establish a direct proportionality between the macroscopic critical current and the microscopic disorder of the lattice from the onset to the maximum of the second peak region.

A vortex (flux-line) lattice is formed in a type-II superconductor by applying a magnetic 
field ($H_\text{a}$) \cite{Abr57a,Bra95a}. Single vortices have a tube-like shape and are  roughly aligned along the field orientation. They have a normal-conducting core with a radius 
in the nanometer range and are surrounded by circulating supercurrents generating a magnetic flux quantum ($\Phi_0 \simeq 2.07 \times 10^{-15}$\, Vs). 
Their density is basically proportional to the magnetic field via $n \simeq \mu_0 H_\text{a} / 
\Phi_0$, where $\mu_0 = 4\pi \times 10^{-7}$\, NA$^{-2}$. The circular currents 
make any two flux-lines repel each other, which results in an ordered 
hexagonal lattice, in which no macroscopic currents over macroscopic ranges can be accumulated. But pinning of flux-lines by material defects opposes 
that ordering and may result  in a disordered lattice. Changing, for instance, the external magnetic field leads to a macroscopic gradient of the vortex density and thus to a macroscopic current when pinning is sufficiently strong \cite{Bra95a}.

\section*{Results and discussion}

\subsection*{Experiment}

\begin{figure}[]
  \centering
    \includegraphics[width = 8.5 cm]{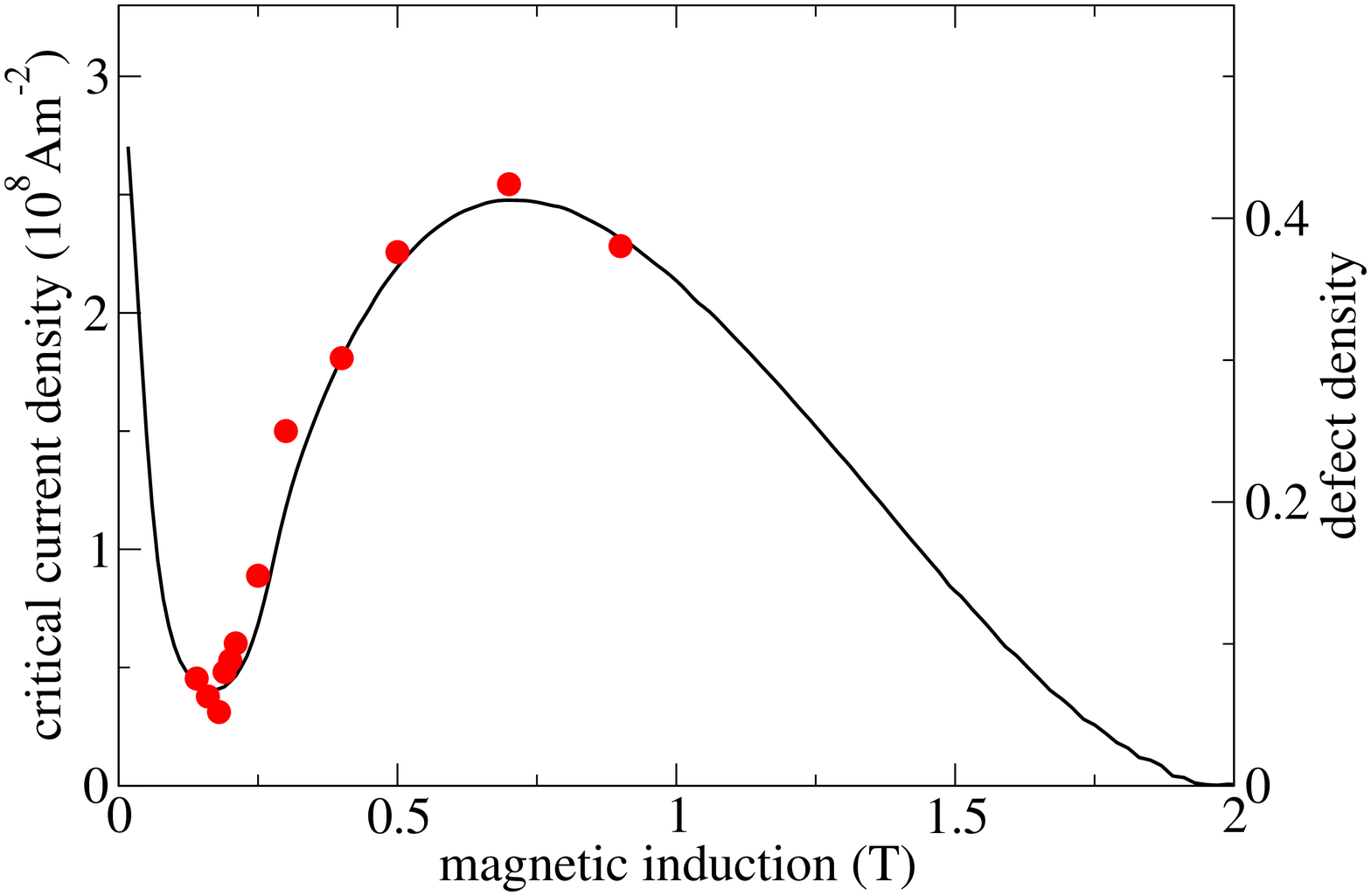}
   \caption{\textbf{Comparison of the critical current density and the defect density.} 
The critical current density (solid line) and the defect density (symbols) of 
the vortex lattice of the irradiated (F$_\text{n}$ = 3 $\times$ 10$^{21}$\,m$^{-2}$) 
NbSe$_2$ single crystal are almost directly proportional from 0.14 to 0.9\,T at 4.2\,K. 
\label{fig1}}
\end{figure}

The experiments were carried out on NbSe$_2$ single crystals (see 
methods). The critical current density of the as-grown (unirradiated) samples had just one maximum at a low magnetic field and declined rapidly with 
increasing field. Introducing material defects, capable of pinning vortices, by 
exposing the samples to neutron irradiation made the current density increase 
and, in addition to the low-field maximum, a second current peak appear at high fields, called the second peak effect (see figure~\ref{fig1} and Supplementary section I). It should be noted that the samples have a finite thickness of some 100\,$\mu$m, along which vortices can be bent, making the vortex matter system more three-dimensional. Double-sided decoration measurements revealed rather stiff vortices in NbSe$_2$ \cite{Mar98a}.

Mapping the density of states of the sample surface, scanning tunneling 
spectroscopy reveals the locations of the vortices (see methods).
To get the same history-dependent vortex matter state as with the macroscopic current measurements, it was necessary to first set the temperature to 4.2\,K and then apply the magnetic field. This procedure leads to a small gradient in the macroscopic vortex density proportional to the critical current, establishing a corresponding driving force on the vortices 
\cite{Bra95a}.  

\begin{figure*}[]
  \centering
  \includegraphics[width = 0.95\textwidth]{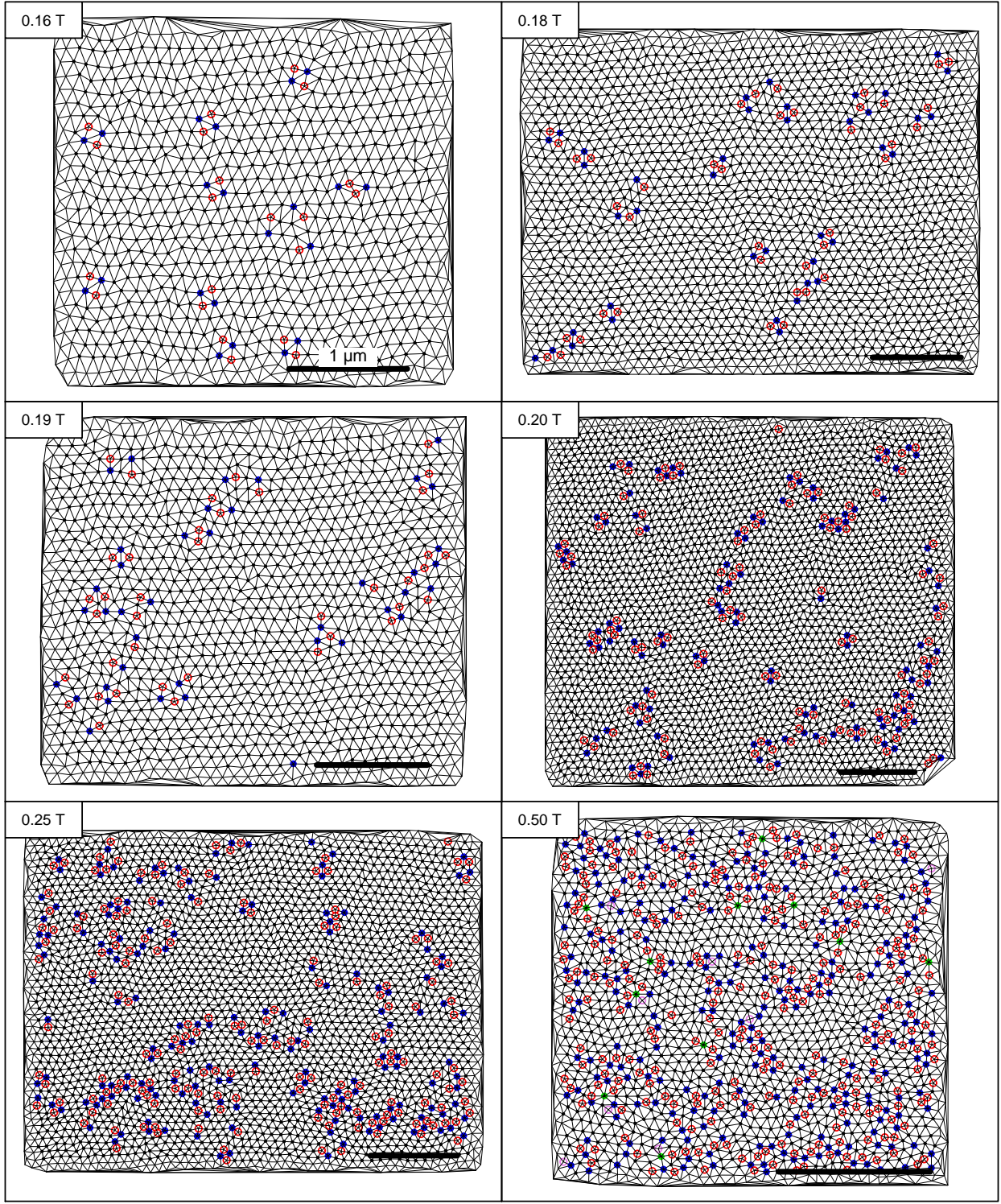}
   \caption{\textbf{The order-disorder transition of the vortex lattice in NbSe$_2$ at 
4.2\,K.} The vortex lattice of NbSe$_2$, shown at 0.16, 0.18, 0.19, 0.20, 0.25, 
and 0.50\,T (cp. with figure~\ref{fig1}), becomes more disordered with increasing field. Each thin line indicates 
a bond between two adjacent vortices, the large (colored) symbols indicate 
lattice defects, namely vortices with four ({\color{violet} $\circ$}), five 
({\color{red} $\circ$}), seven ({\color{blue} $\bullet$}), or eight 
({\color{green} $\bullet$}) nearest neighbors. The horizontal bars correspond to a length of 1\,$\mu$m. The image boundaries were not used for the evaluation. \label{fig2}}
\end{figure*}

\subsection*{The vortex lattice and the critical current density}

Figure~\ref{fig2} shows the vortex lattice of the sample of figure~\ref{fig1} at 
several magnetic fields from 0.16\,T, which is close to the minimum of the 
current, to 0.50\,T, which is deep in the second peak region. Here
each line connects two adjacent vortices, hence each intersection marks the 
position of a vortex. As each vortex has six nearest neighbors in a
hexagonal lattice, a lattice defect is defined as a vortex having more or 
less than six neighbors and is highlighted by a large colored symbol in the figures.

Figure~\ref{fig2} makes it manifest that the disorder, reflected 
by the lattice defect density, increases with field. Figure~\ref{fig1} shows that 
the lattice defect density is almost directly proportional 
to the macroscopic critical current density from about 0.14 to 0.9\,T.
Such a relation is not a-priori expected and was indeed not observed at higher fields.  The 
microscopic vortex lattice and the macroscopic currents were determined in the same sample.

\subsection*{Defects in the vortex lattice and the order-disorder transition}

Different kinds of defect \cite{Gru07a} appear in the two-dimensional 
vortex lattice images. A single defect, having no defect as a direct neighbor, is called 
a disclination. Two adjacent defects from which one has five and the other seven neighbors (shown by a red-blue pair in the figures) indicate a dislocation. Two connected dislocations form a dislocation pair. Finally, also larger defect clusters show up.

The most established theoretical explanation of how a two-dimensional particle 
system undergoes an order-disorder transition was worked out by Kosterlitz, 
Thouless, Halperin, Nelson, and Young (KTHNY \cite{Kos73a,Nel79a,Hal78a,You79a}). In the ordered state, characterized by a quasi-long-range translational and a long-range orientational order, defects appear only in the form of dislocation pairs. Pushing the 
system towards the disordered state first generates the so-called hexatic state by dissolving the pairs into single dislocations, making the translational order short-range. The final disordered state is dominated by disclinations, making also the orientational order short-range. Alternatively, Chui \cite{Chu83a} proposed that 
the dislocations form grain boundaries instead of dissolving into disclinations. Both scenarios were backed by experiments and simulations.

Previous measurements  \cite{Hec14a} revealed a rather well ordered defect-free vortex lattice in an unirradiated clean NbSe$_2$ single crystal at fields where the critical current vanished. Magnetic decoration images by Kim et al.\cite{Kim99a} showed that defect-free regions may contain more than 10$^5$ vortices in clean superconductors. In contrast, several dislocation pairs and a small number of single dislocations showed up in a sample irradiated to a low neutron fluence (F$_\text{n}$ = 10$^{21}$\,m$^{-2}$) at fields between the two peaks where the critical current was also close to zero (Supplementary section II).

\subsection*{The different stages of the order-disorder transition}

\begin{figure}[]
  \centering
    \includegraphics[width = 8.5 cm]{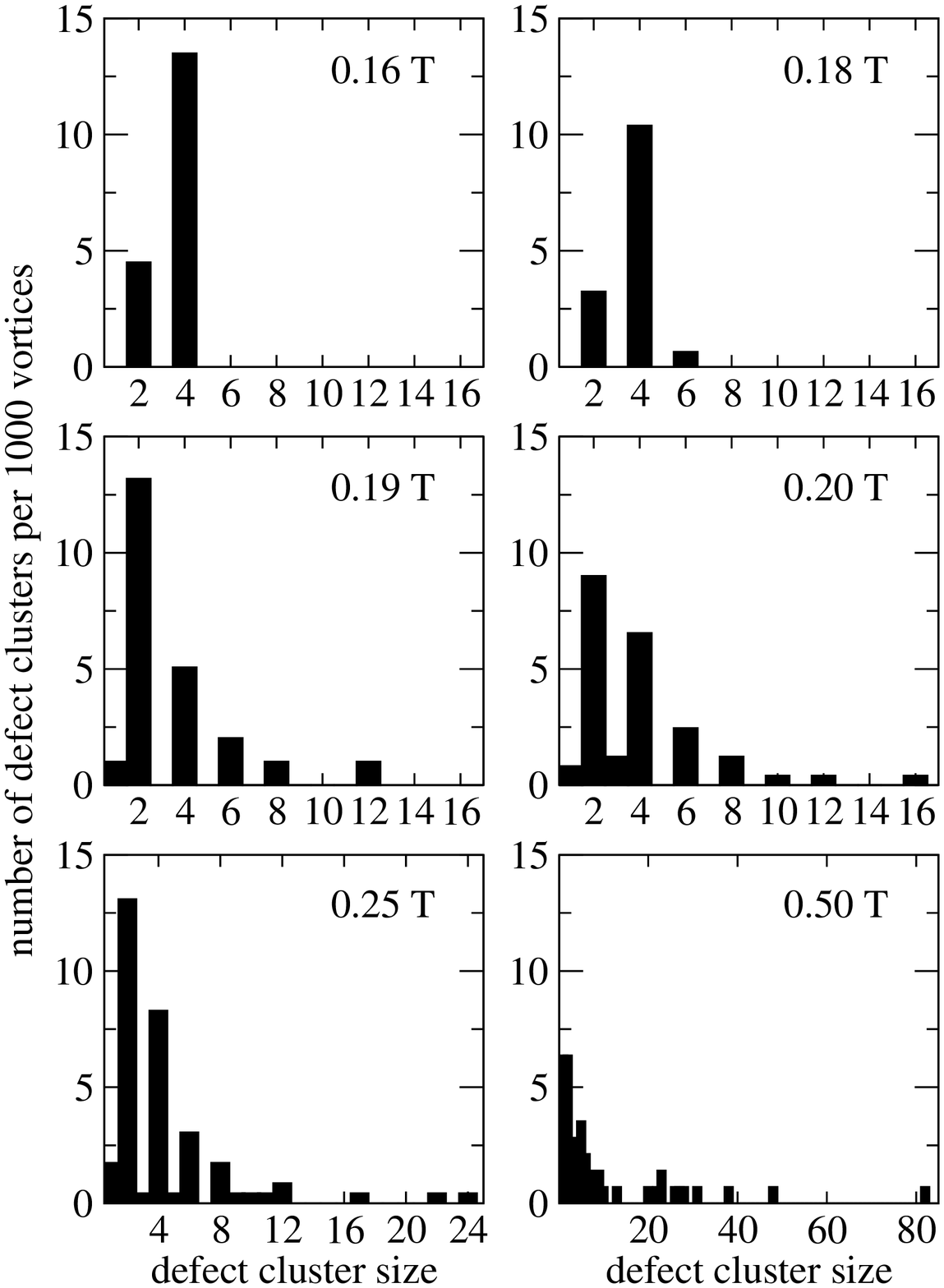}
   \caption{\textbf{Histograms of the defect cluster size.} The panels show the number of 
defect clusters per 1000 vortices as a function of the cluster size of all 
lattices of figure~\ref{fig2}. \label{fig3}}
\end{figure}

Now I come to the details of figures~\ref{fig2} and ~\ref{fig3}, which refer to 
a higher irradiated (F$_\text{n}$ = 3 $\times$ 10$^{21}$\,m$^{-2}$) sample, whose 
current, shown in figure~\ref{fig1}, does not vanish between the peaks. At 
0.16\,T, a field close to the minimum of the critical current, several dislocation pairs and a few single dislocations appear. To compare lattices with 
different sizes, I normalized the number of defects to a lattice with 1000 
vortices. This gives 13.5 dislocation pairs and 4.5 single dislocations 
(figure~\ref{fig3}). No other defects are visible. Note that all dislocation 
pairs form loops with alternating coordination numbers between adjacent 
defects, thus minimizing the Burgers vector \cite{Gru07a} and the corresponding defect energy, and destroying the order merely locally. It is reasonable that dislocation pairs dominate at low 
disorder, for they can easily be created just by locally displacing the involved 
vortices. In contrast, single dislocations are topological defects, which cannot 
be created by such simple transformations  \cite{Gru07a}, but two are obtained by splitting a 
dislocation pair, or they are introduced at the sample edges. Likewise, dissolving a dislocation leads to two disclinations. The pairs also dominate in the lower irradiated sample (F$_\text{n}$ = 10$^{21}$\,m$^{-2}$) at fields between the two current peaks and also form loops with alternating coordination number (Supplementary section II), but even here single dislocations exist. Some of these single dislocations may be remnants of the more disordered state at low fields, where the first current peak appears. Note that some vortices may be pinned strongly enough to maintain their position during the magnetization process.

Increasing the field to 0.18\,T leads to merely minor modifications. There are still more 
pairs (10.4) than single dislocations (3.2), but, for the first time, also a 
larger defect cluster, consisting of six vortices, emerges. Moreover, some of 
the pairs no longer exist as loops but form lines that are not closed, leading to rather
large Burgers vectors and defect energies. This may be the first stage of
a process that dissolves the pairs into single dislocations. It also becomes apparent that the defects are not homogeneously distributed in the lattice but rather form chains.

At 0.19\,T the changes are more fundamental. In contrast to the low-field 
results, the number of single dislocations (13) exceeds that of the pairs (5) (figure~\ref{fig3}). Most single dislocations are located close to another one, 
indicating that they have just been separated. Additionally, larger clusters 
with up to twelve vortices and one disclination, located, however, at the image 
boundary and hence presumably being part of a larger cluster, show up. Apart from this 
disclination, all clusters consist of an even number of vortices. 

With increasing field, the basic trends, observed at 0.19\,T, carry on. 
The number of single dislocations and pairs does not 
change considerably, while the size of the larger clusters grows. The number of 
clusters decays roughly exponentially with their size (figure~\ref{fig3}). 
Disclinations emerge more frequently though do not become dominant.

Going to 0.2\,T slightly increases the critical current, though the field is still close to the onset of the second peak regime. Here, the formation of grain boundaries, consisting of defect clusters, becomes manifest. At 0.25\,T the critical current has increased by more than a factor of two and the disorder has grown accordingly. The defect cluster chains become more bulky though defect-free grains are still visible.

Finally, at 0.5\,T, which is deep in the second peak region, the lattice is quite disordered. The defect density is about one third, and for the first time, vortices with more than seven or less than five nearest neighbors emerge. The defect amalgamations are no longer line-shaped but 
rather bulky, yet defect-free vortex grains are still present. A further increase of the magnetic field to 1.6\,T basically raises the disorder but does not alter the nature of the vortex lattice structure.

Although the finite sample thickness leads to a more three-dimensional character of the vortex lines, some similarities with KTHNY theory are striking. In particular I observed in accordance with KTHNY theory that in the most ordered lattice, dislocation pairs dominate, which first form closed loops, so that the order is  only locally destroyed, and later also become elongated. Then, single dislocations, destroying the quasi-long-range translational order, appear more frequently. But finally, in contrast to KTHNY theory, which predicts disclocations to dissolve into disclinations, though in agreement with Chui's proposal \cite{Chu83a}, the dislocations become clustered and form a grain boundary structure getting more bulky with increasing disorder. 

\subsection*{Correlation functions}

\begin{figure}[]
  \centering
    \includegraphics[width = 8.5 cm]{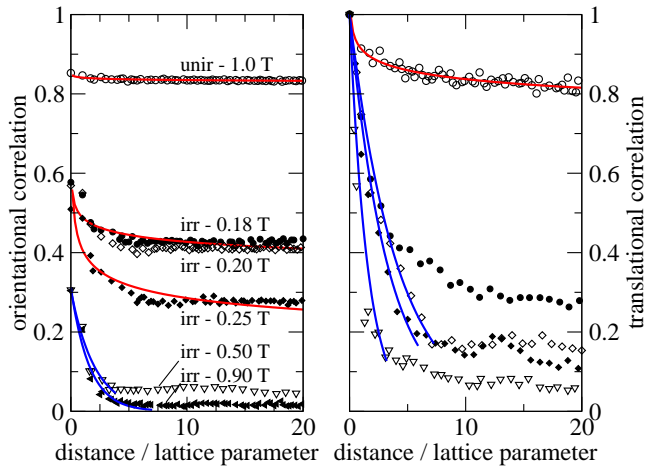}
   \caption{\textbf{The correlation functions of the vortex lattices.} The symbols show 
the (upper) envelope of the orientational (left panel) and of the translational 
(right panel) correlation functions of the unirradiated sample at 1\,T 
 and of the irradiated sample of figure~\ref{fig2} at different 
magnetic fields. The fit functions (solid lines) follow either a power law (red) 
or an exponential (blue) decay behavior. \label{fig4}}
\end{figure}

Correlation functions offer additional insight into the ordering state of our 
systems (see methods). The translational correlation function measures the 
translational symmetry of the system and the bond-orientational correlation 
function the relative orientation of two bonds between nearest neighbors as a 
function of distance. Both functions change from one for a perfectly ordered to 
zero for a completely disordered state. 

Figure~\ref{fig4} shows these functions for our systems. The highest order is 
found in the defect-free lattice of the unirradiated sample. Its slightly 
decaying though quasi-long-range translational order and its virtually constant 
orientational order agree with the properties of a Bragg glass \cite{Gia95a,Kle01a}. 
In the disordered lattices, that is, at and above 0.5\,T, both functions decay 
exponentially for not too large distances and have similar correlation lengths 
of 1-2 lattice parameters. This would match the vortex glass behavior, though 
the lattice at 0.5\,T is not amorphous. The flat long-range behavior
reflects the fact that the mean orientation of different vortex grains does not 
change significantly, presumably an aftermath of the ordered state that had been run through during applying the field at a constant temperature. Additionally, the finite sample thickness may also contribute to the unusual long-range behavior. At higher fields the translational correlation does not change significantly, while the orientational correlation further declines.

Near the onset of the second peak, for instance when the field changes from 0.18 to 0.20\,T, 
the orientational order remains almost unchanged and is quasi long-range, whereas the translational order decreases appreciably and has a much shorter correlation length in both cases. This is what one expects of the hexatic state \cite{Nel79a}. The observation of the hexatic state, associated with the appearance of free dislocations, was also reported for the melting process of the vortex lattice in a superconducting tungsten-based thin film \cite{Gui09a}. The hexatic phase in a three-dimensional vortex lattice was discussed in Ref.~\onlinecite{Mar90a}.

\subsection*{Further discussion}

In conclusion, the order-disorder transition of a superconductor's flux-line lattice follows reasonably well the theoretical predictions of the grain boundary scenario. While in most cases, the systems are cooled down from a disordered state, our system reached the final state by increasing the magnetic field at a constant temperature and thus always first passed the low-field disordered state and then the most ordered state at the onset of the second peak. But only this procedure makes a comparison of the microscopic vortex lattice with the macroscopic critical current useful. Moreover, it is a change in temperature that evokes the transition in most experiments, whereas it is a change in magnetic field in our case. Increasing the magnetic field increases the number of vortices and thus reduces the distances between them, as it is the case when pressure is applied to a conventional system. 

The experiments confirm that the transition is induced mainly by pinning and not by thermal effects, since no high-field disordered state, at least below 1.7\,T, which is 85 per cent of the upper critical field, was observed in the unirradiated sample, in which vortex pinning is weak. Moreover, the disordered vortex lattice configurations did not change significantly with time, as was found from measuring the same sample area several times in immediate succession. This agrees with the common assumption that the order-disorder transition is a solid-to-solid transition \cite{Mik01a,Kie04a}, at least in low-temperature superconductors. It was shown theoretically \cite{Mik01a,Kie04a} that in samples with a very small Ginzburg number, such as in NbSe$_2$ (10$^{-5}$ - 10$^{-6}$), thermal effects are not expected to become relevant at fields lower than some 90 per cent of the upper critical field, which is far from our transition field. Pinning effects extend the region where thermal energy matters, but the predicted shift of the melting line was found to be small for NbSe$_2$ even when pinning effects are strong \cite{Mik03a}.

The assumption of a two-stage nature of the order-disorder transition in vortex matter was recently confirmed by two other scanning tunneling spectroscopy studies \cite{Gui14a,Gan14a}. In the first, Guillamon et al.\cite{Gui14a} investigated a tungsten-based superconducting thin film, in which the order-disorder transition was generated by a modulation of the sample thickness. No lattice defects were found at low magnetic fields. In the first stage of the transition, pair and single dislocations showed up, and in the second stage, the dislocation density grew strongly and free disclinations were created. The translational and orientational correlation functions revealed good agreement with KTHNY theory.

Ganguli et al. \cite{Gan14a} analyzed a cobalt-doped NbSe$_2$ single crystal having a second peak effect at high magnetic fields. They also observed a proliferation of dislocations in the first stage of the transition and the appearance of single disclinations, driving the lattice into an amorphous glass, in the second stage. They showed that the increasing and the decreasing field branch led to similar vortex structures, but differences were observed in the long-range behavior of the correlation functions at fields in the transition region, which confirms the assumption suggested above that the deviations of the correlation functions from KTHNY theory for long distances may be explained by the specific history of the magnetization process. Cooling the sample at a constant field led to the expected higher disorder in the lattices.

Though all three studies confirm the two-step nature of the transition, a detailed comparison reveals some remarkable differences, which may be caused by different material properties and sample thicknesses, leading to a more three- (Ref.~\onlinecite{Gan14a} and this work) or a more two-dimensional (Ref.~\onlinecite{Gui14a}) character of the vortex matter. First, in the present work, I found the defect structure of the rather ordered lattice to be dominated by dislocation pairs, which agrees with the prediction of KTHNY theory. In contrast, merely single dislocations showed up in Ref.~\onlinecite{Gan14a} at this stage. Second, I found the defects to form a grain boundary structure, getting more bulky with increasing field. Such a structure was not reported by the other groups, though several vortex lattice images of Ref.~\onlinecite{Gan14a} might indicate an inhomogeneous defect distribution.

Vortex lattice defects concentrated mainly in grain boundaries were reported by Fasano et al. \cite{Fas02a,Fas08a} in NbSe$_2$. These results were, however, obtained at very low magnetic fields, namely at 3.6\,mT, which was not in the second peak regime but close to the first critical current peak. Employing magnetic decoration, they were able to assess large sample areas and thus to detect vortex grains including several hundred vortices in both field-cooled and zero-field-cooled experiments. Based on the assumption that the field-cooled lattices had been frozen in close to the upper critical field in the second peak region, they concluded that the vortex lattice disorder does not correlate with the macroscopic current in the second peak region, which is in contrast to the proportionality found in this work. However, the proportionality between current and lattice disorder was only shown for fields between the onset and the maximum of this region in this work, while the frozen lattice of Ref.~\onlinecite{Fas02a} would refer to the high-field or high-temperature end of this region, where the correlation can be quite different. A grain boundary structure was also found in  simulations \cite{Mig03a,Das03a,Cha04a,Mor05a,Mor09a} of the two-dimensional vortex matter. Other two- or quasi-two-dimensional systems such as electron sheets \cite{Gri79a}, dusty plasmas \cite{Tho96a,Nos09a}, colloidal \cite{Mur87a,Mar96a,Deu13a} and other systems \cite{Str88a,Gru07a} were found to become disordered by the same or a similar two-stage process, though only some of these systems developed a grain-boundary structure.

\section*{Methods} 

\subsection*{Samples} The material used for this study is NbSe$_2$, which becomes
superconducting below some 7.2\,K. All experiments were carried 
out at 4.2\,K with the magnetic field parallel to the uniaxial axis of the single crystal's hexagonal unit cell, for which the upper critical field is some 2\,T. 
Recently, NbSe$_2$ was shown to be a two-band superconductor similar to 
MgB$_2$ \cite{Zeh10a}. Our samples had a size of 1-3\,mm in lateral direction and 100 - 
200\,$\mu$m parallel to the uniaxial axis.

\subsection*{Neutron irradiation} Neutron irradiation of the samples took place 
in a TRIGA MARK II research fission reactor \cite{Web86a}. This procedure introduces crystal defects that are  efficient for vortex pinning 
in many superconductors. For instance, in MgB$_2$ \cite{Mar08a} and YBa$_2$Cu$_3$O$_{7-\delta}$ \cite{Fri94a}, transmission electron microscopy revealed spherical defects with a diameter close to the superconducting coherence length. Neutron irradiation tremendously strengthens vortex pinning also in NbSe$_2$, as shown by
its effects on the critical current (Supplementary section I). One sample was irradiated to a fast neutron ($E > 0.1$\,MeV) fluence of 1 $\times$ 10$^{21}$\,m$^{-2}$ and the sample to which most results of this article refer to 3 $\times$ 10$^{21}$\,m$^{-2}$. 

\subsection*{Critical current density} The critical current density was obtained 
from SQUID magnetometry by measuring the magnetic moment as a function of  
magnetic field at 4.2\,K. Assuming the current to be constant in the sample, one can derive the 
critical current from the hysteresis of the magnetic moment in 
increasing and decreasing field \cite{Zeh09a}. The unirradiated samples showed 
a finite though small critical current only at low fields. A low neutron fluence 
enhanced the current significantly at low fields and additionally produced a 
second peak close to the upper critical field. With increasing fluence the 
low-field current became larger and the second peak more prominent.

\subsection*{Scanning tunneling spectroscopy}
Scanning tunneling spectroscopy reveals the electronic density of states at the 
sample surface. In a superconductor the procedure produces 
the well-known energy gap of the quasi-particles near the Fermi level. This gap, 
however, does not exist in the vortex core, and hence the tunneling signal changes, 
thus identifying the vortex core positions, which makes scanning tunneling spectroscopy the sole known method for measuring the vortex lattice in real-space at arbitrary magnetic fields \cite{Hes89a,Iav08a,Han12a,Hec14a,Fis07a,Sud14a}. 

For the measurements, a sample was cleaved in air and then put into the sample tube, which was first evacuated and then flooded with pure helium 
gas. The measurements were carried out with a PtIr tip. Having placed the sample in a helium cryostat, I cooled it to 4.2\,K and then applied a magnetic 
field. Then, after waiting for several hours, the surface topography was mapped by a constant-current mode measurement and, at the same time, the density of states at an energy level slightly 
above the Fermi level but below the energy gap width of some 2\,meV by a lock-in 
technique (Supplementary section III). The densities of state maps usually showed a clear 
contrast for the vortex core positions (Supplementary section III). Applying simple filters such as a 
Gaussian filter to the raw data reduced the noise, so that the vortex core 
positions were easily acquired by seeking all local maxima. The nearest
neighbors of each vortex were determined by a Delaunay triangulation.

\subsection*{Correlation functions}
The translational correlation function \cite{Nel79a} is defined as
\begin{equation}
  G_\text{T} (|\vec{r} - \vec{r}'|) = \langle e^{i \vec{G} \vec{u}}  e^{-i 
\vec{G} \vec{u}'} \rangle
\end{equation}
where $\vec{G}$ is a reciprocal lattice vector, and $\vec{u}$ is the 
displacement of a vortex at $\vec{r}$ from its lattice position $\vec{R}$, that 
is, $\vec{u}$ = $\vec{r} - \vec{R}$ and $e^{i \vec{G} \vec{R}} = 1$. The brackets 
indicate an averaging over all vortex pairs. The reciprocal lattice vector was 
determined by seeking the vector with the largest value near the boundary of the first Brillouin zone after a Fourier transformation of the lattice. The 
distances were discretized into interval lengths of $0.1\,a_0$.

For the bond orientational correlation function \cite{Nel79a}, one has to calculate
\begin{equation}
  G_6 (|\vec{r} - \vec{r}'|) = \langle b_6(\vec{r})~ b_6^\star  
(\vec{r}')\rangle
\end{equation} with
\begin{equation}
  b_6(\vec{r}) = \frac{1}{n}\sum_{j=1}^n e^{i6\phi(\vec{r},\vec{r}_\text{j})}.
\end{equation}
The sum runs over all nearest neighbors j of the vortex at $\vec{r}$, and 
$\phi(\vec{r},\vec{r}_\text{j})$ is the angle of the bond connecting the two 
vortices with respect to the $x$-axis.

\begin{acknowledgments}
I would like to thank F. Sauerzopf, J. Hecher, M. Eisterer, and H.W. Weber for supporting this work and discussing the results with me, and P.H. Kes for providing the samples.
This work was supported by the Austrian Science Fund under Contract No. 21194
and 23996.
\end{acknowledgments}

\section*{Author contributions}

The author performed experiments, analysed data and wrote the paper.

\section*{Competing financial interests}

The author declares no competing financial interests.

\end{document}